\documentclass[twocolumn,secnumarabic,amssymb,amsmath,superscriptaddress,nofootinbib,tightenlines,nobibnotes,aps,prc]{revtex4-2}
\usepackage[normalem]{ulem}
\usepackage{graphicx,xcolor}
\usepackage[unicode]{hyperref}

\begin{document}


\title{Charged-current quasielastic-like neutrino scattering from $^{12}$C in the coherent density fluctuation model with two-nucleon emission}

\author{M.V.~Ivanov\footnote{martin.inrne@gmail.com}}
\affiliation{Institute for Nuclear Research and Nuclear Energy, Bulgarian Academy of Sciences, Sofia 1784, Bulgaria}
\affiliation{Department of Mathematics and Physics, Faculty of Mathematics and Natural Sciences, South-West University ``Neofit Rilski'', Blagoevgrad, Bulgaria}
\author{A.N.~Antonov}
\affiliation{Institute for Nuclear Research and Nuclear Energy, Bulgarian Academy of Sciences, Sofia 1784, Bulgaria}

\date{\today}

\begin{abstract}

The quasielastic cross-sections of charged-current neutrino and antineutrino scattering on $^{12}$C are calculated using the coherent density fluctuation model with a relativistic effective mass $m_N^* =0.8 m_N$ (CDFM$_{M^*}$). The model explicitly considers the modification of the relativistic effective mass of the nucleon within the relativistic mean field (RMF) model of nuclear matter. In addition, our calculations include neutrino-induced two-particle emission processes, which are evaluated within the RMF model of nuclear matter. Utilizing the CDFM$_{M^*}$, we provide predictions for the neutrino and antineutrino cross sections of $^{12}$C, which have been observed in accelerator experiments, such as MiniBooNE, T2K, and MINERvA. Also, we analyze the axial form factor value for the excitation of the $\Delta$ at zero momentum transfer (commonly denoted as $C^A_5 (0)$) which is important for the treatment of the $\Delta$ current in the meson-exchange currents (MEC) calculation. In addition, the quasielastic results obtained within CDFM$_{M^*}$ model are thoroughly evaluated for different regions of the momentum transfer.

\end{abstract}

\maketitle

\section{Introduction\label{sec:introduction}}

In recent years, a comprehensive experimental program has been developed, focusing on accelerator-based neutrino oscillation experiments. The primary goal of this program is to improve our understanding of neutrino properties by precisely measuring the oscillation parameters and examining the weak CP-violating phase. The proper understanding of the interactions between neutrinos and atomic nuclei is crucial in these experiments for achieving reliable measurements of neutrino oscillation parameters. Nevertheless, modeling lepton-nucleus interactions presents challenges due to the intricacies of nuclear effects. Approaches like the relativistic Fermi gas (RFG) model offer a basic description of the quasielastic (QE) cross section, yet they fall short in accounting for significant aspects such as medium modifications, finite-size effects, nuclear correlations, and final-state interactions. Consequently, more sophisticated nuclear models are necessary to provide an accurate and dependable representation of the data.

The phenomenon of superscaling is based on the scaling characteristics of electron scattering data and was first examined within the context of the RFG model~\cite{PhysRevC.38.1801, Barbaro1998137, PhysRevLett.82.3212, PhysRevC.60.065502, PhysRevC.65.025502, BCD+04}. At sufficiently high momentum transfer, the inclusive differential $(e, e')$ cross-sections, divided by a suitable function that accounts for the single-nucleon content of the problem, depend on one kinematical variable, known as the scaling $\psi$-variable (this behavior is called scaling of the first kind). When the resulting function is approximately uniform across all nuclei, it is termed scaling of the second kind. When both types of scaling are applicable, the cross-section exhibits superscaling. It was observed in Ref.~\cite{PhysRevC.60.065502} that the experimental data have a superscaling behavior for large negative values of $\psi$ (up to $\psi\approx -2$), while the predictions of the RFG model are $f(\psi)=0$ for $\psi \leq -1$. This leads to the necessity of considering superscaling in realistic finite systems. One model to achieve this was developed in Refs.~\cite{A7, A8} within the coherent density fluctuation model (CDFM)~\cite{anton1, anton2, antonov_bjp_1979, antonov_nucleon_1980, antonov_spectral_1982, antonov_extreme_1985, antonov_natural_1989, PhysRevC.50.164}, which is related to the $\delta$-function limit of the generator coordinate method \cite{A7, PhysRev.108.311}. It has been demonstrated in~\cite{A7, A8, A9} that superscaling in nuclei can be quantitatively elucidated based on the analogous behavior of the high-momentum components of the nucleon momentum distribution across light, medium, and heavy nuclei. It is widely recognized that this phenomenon is associated with the effects of nucleon-nucleon (\emph{NN}) correlations in nuclei (see, \emph{e.g.}~\cite{anton1, anton2}).

In our previous works~\cite{A7, A8, A9, A10} we obtained the CDFM scaling function $f(\psi)$ starting from the RFG model scaling function $f_\text{RFG}(\psi)$ and convoluting it with the weight function $|F(x)|^2$ that is related equivalently to either the density $\rho(r)$ or the nucleon momentum distribution $n(k)$ in nuclei. Thus, the CDFM scaling function is an infinite superposition of weighted RFG scaling functions. This approach improves upon RFG and enables one to describe the scaling function for realistic finite nuclear systems. The CDFM scaling function has been used to predict cross sections for several processes, such as the inclusive electron scattering in the QE and $\Delta$- regions~\cite{A10, A13} and neutrino (antineutrino) scattering both for charge-changing (CC)~\cite{A13} and for neutral-current (NC)~\cite{A12} processes. In our work~\cite{A10} we reproduce experimental data of the inclusive electron scattering in the QE-region using the CDFM scaling function which is obtained using a suitable parameterization of the RFG scaling function and by the coefficient $c_1$, which helps us to account for the experimental fact of the asymmetry of the scaling function. The value of the coefficient $c_1$ ($c_1 \neq 3/4$) is taken in accordance with the empirical data ($c_1$ depends on the value of the momentum transfer in the QE peak).

The superscaling analysis with relativistic effective mass (SuSAM), a variant of superscaling analysis (SuSA) approach designed to improve the description of quasielastic $(e, e')$ data by incorporating nuclear dynamics inspired by the Relativistic Mean Field (RMF) theory of nuclear matter, is developed in Refs.~\cite{Mar17, Ama18}. This approach embeds medium effects from the outset and preserves gauge invariance. In Ref.~\cite{1njz-1m49} an improved scaling analysis, SuSAM-v2, is established. The model builds upon the SuSAM framework by preserving its key dynamical ingredient, the relativistic effective mass, while introducing a more constrained and controlled extraction of the scaling functions. The resulting SuSAM-v2 model simultaneously describes inclusive quasielastic cross sections and both longitudinal and transverse response functions in electron scattering. Also, the SuSAM-v2 model shows an improved prediction compared to the previous SuSAM to neutrino-nucleus scattering.

In our recently published studies~\cite{PhysRevC.109.064621, universe11040119}, we established and explored a novel scaling approach known as CDFM$_{M^*}$, which incorporates an effective mass $m_N^*$ and defines $M^* = m_N^*/m_N$. This approach applies the scaling function derived from the CDFM, which is fundamentally based on the scaling function of the RFG. Additionally, we used a new scaling variable $\psi^*$, which is derived from the scaling characteristics of the RMF model in nuclear matter~\cite{Ama15}. As it is shown in Refs.~\cite{Ama15, Ama17, Mar17, Ama18, Rui18} the newly defined scaling function $f^*(\psi^*)$ integrates dynamical relativistic effects through the incorporation of an effective mass. Within the framework of the CDFM$_{M^*}$ model, we derived a scaling function $f^\text{QE}(\psi^*)$ applicable in the quasielastic (QE) region, utilizing the empirical density distribution of protons to ascertain the weight function $|F(x)|^{2}$.
The results, using the new scaling function $f^\text{QE}(\psi^*)$, indicate that the CDFM$_{M^*}$ model enhances the transverse components of the electromagnetic current. This observation supports the conclusion that an effective nucleon mass reduction to $M^* = 0.8$ results in an increased transverse response, consistent with the RMF model's inclusion of dynamical relativistic effects, such as the enhancement of the transverse response attributed to the lower components of nucleon spinors. Furthermore, we calculated the inclusive $(e,e')$ and (anti)neutrino differential quasielastic cross section utilizing the scaling function $f^\text{QE}(\psi^*)$. This analysis incorporated the two-particle--two-hole (2p--2h) MEC contribution derived from the RFG model. The findings indicate a satisfactory representation of the electron data and the (anti)neutrino data from MiniBooNE, MINERvA, and T2K experiments.

In the present work we use the CDFM$_{M^*}$ approach to calculate (anti)neutrino CC quasielastic differential cross sections and compare the results with the data from MiniBooNE~\cite{PhysRevD.81.092005, miniboone-ant}, T2K~\cite{PhysRevD.93.112012} and MINERvA~\cite{PhysRevD.99.012004, PhysRevD.97.052002} experiments. In our work, the model also includes contributions from 2p--2h excitations computed within the RMF model of nuclear matter~\cite{PhysRevD.104.113006, PhysRevD.108.013007, PhysRevD.108.113006}, in contrast to those obtained using the RFG model in~\cite{PhysRevC.109.064621, universe11040119}. We analyze the axial form factor value for the excitation of the $\Delta$ at zero momentum transfer, $C^A_5 (0)$, which is important for the treatment of the $\Delta$ current in the MEC calculation. In addition, the quasielastic results obtained within CDFM$_{M^*}$ model are thoroughly evaluated for different regions of the momentum transfer. Section~\ref{sec:scheme} contains a brief review of the theoretical scheme to obtain the CDFM$_{M^*}$ scaling function as well as an overview of MEC in order to apply them to the calculations of the (anti)neutrino CC quasielastic differential cross sections. Our main results for the latter are presented in  Section~\ref{sec:results}. Finally, in Section~\ref{sec:conclusions}, we draw our conclusions of the present work.

\section{Theoretical scheme\label{sec:scheme}}

\subsection[]{Cross section and scaling function in CDFM$_{M^*}$}

In this Section we present the main ingredients of our calculations of QE cross sections of charged-current neutrino and antineutrino scattering on $^{12}$C.

The double-differential cross section has the form (see Refs.~\cite{Ama05a, Ama05b,PhysRevC.109.064621}):
\begin{multline}
\frac{d^2\sigma}{dT_\mu d\cos\theta_\mu} = \sigma_0 \big[ V_{CC} R_{CC} + 2{V}_{CL} R_{CL} + {V}_{LL} R_{LL} +\\
+ {V}_{T} R_{T} \pm 2{V}_{T'} R_{T'} \big] \, ,\label{nuxsec}
\end{multline}
with
\begin{equation}
\sigma_0= \frac{G^2\cos^2\theta_c}{4\pi} \frac{k'}{\epsilon}v_0.\label{nuxsec0}
\end{equation}

In Eqs.~(\ref{nuxsec}) and~(\ref{nuxsec0}) the energies of the incident neutrino (antineutrino) and detected muon are $\epsilon=E_{\nu}$, $\epsilon'=m_\mu+T_\mu$ with corresponding momenta ${\bf k}$ and ${\bf k}'$. The four-momentum transfer is $k^\mu-k'{}^\mu=(\omega,{\bf q})$ and $Q^2=q^2-\omega^2 > 0$. $\theta_\mu$ is the scattering angle. In Eq.~(\ref{nuxsec0}) $G=1.166\times 10^{-11}\quad\rm MeV^{-2} \approx 10^{-5}/ m_p^2$ is the Fermi constant, $\theta_c$ is the Cabibbo angle, $\cos\theta_c=0.975$, and $v_0= (\epsilon+\epsilon')^2-q^2$ is the kinematic factor. $R_K(q,\omega)$ are five nuclear response functions with indices: $CC$ -- charge-charge, $CL$ -- charge-longitudinal, $LL$ -- longitudinal-longitudinal, $T$ -- transverse, and the function $R_{T'}$ is added for neutrinos ($+$) and subtracted ($-$) for antineutrinos. The coefficients $V_K$ depend only on the lepton kinematics, being independent on the details of the nuclear target.

The response functions $R_K$ are proportional to the corresponding single-nucleon response functions $U_K$ multiplied by the scaling function $f^\text{QE}(\psi^*)$:
\begin{equation} \label{cdfmm*}
R_K = \frac{{\cal N} \xi^*_F}{m^*_N \eta_F^{*3} \kappa^*}  U_K  f^\text{QE}(\psi^*).
\end{equation}
The kinematic coefficients $V_K$, the single-nucleon response function $U_K$, as well as the definition and more details for dimensionless quantities, such as $\eta_F^*$, $\xi^*_F$, $\lambda^*$, $\kappa^*$, $\tau^*$ are given in Ref.~\cite{Rui18}.

The scaling function in the CDFM has been derived in Ref.~\cite{A8}. In the case when the density distribution $\rho(r)$ is used it has the form:
\begin{equation}
f^\text{QE}(\psi^*)= \int_{0}\limits^{\infty}  |F(x)|^{2} \,f_\text{RFG}^\text{QE}[\psi^*(x)]\,{\mathop{}\!\mathrm{d}} x, \label{eq:1}
\end{equation}
in which the CDFM weight function $|F(x)|^2$ can be obtained by means of the density distribution (see for details, e.g. Refs.~\cite{anton1, anton2, antonov_bjp_1979, antonov_nucleon_1980,antonov_spectral_1982}):
\begin{equation}
|F(x)|^{2}=-\frac{1}{\rho_{0}(x)} \left. \frac{d\rho(r)}{dr}\right |_{r=x}, \label{eq:2}
\end{equation}
where
\begin{equation}
\rho_{0}(x)=\frac{3A}{4\pi x^{3}}.\label{eq:3}
\end{equation}
In Eq.~(\ref{eq:1}) the function $f_\text{RFG}^\text{QE}[\psi^*(x)]$ (where $\psi^*(x)=\dfrac{k_{F}x\psi^*}{\alpha}$ depends on $x$ and both energy and momentum transfer through the variable $\psi^*$, see Ref.~\cite{Rui18}) is the scaling function related to the RFG model:
\begin{equation}
f_{\rm RFG}(\psi^*) = \frac{3}{4}(1-\psi^*{}^2) \theta(1-\psi^*{}^2),
\label{rfg*}
\end{equation}
which in the CDFM (Ref.~\cite{A8}) has the form
\begin{equation}
f_\text{RFG}^\text{QE}[\psi^*(x)] = \displaystyle \frac{3}{4} \left[\! 1\!-\!\left( \frac{k_F x \psi^*}{\alpha} \right)^{2}\!\right]
\theta \left[\!1-\!\left(\dfrac{k_F x\psi^*}{\alpha}\right)^2\!\right]. \label{eq:4}
\end{equation}
In Eq.~(\ref{eq:4})
$$\alpha=(9\pi A/8)^{1/3}\simeq 1.52A^{1/3}$$
and the Fermi momentum $k_{F}$ is not a free parameter as it is in the RFG model and can be calculated within the CDFM (or CDFM$_{M^*}$) for each nucleus using the relationship:
\begin{multline}
k_F= \int\limits_{0}^{\infty} |F(x)|^2 \,k_{x}(x)\, {\mathop{}\!\mathrm{d}} x=  \int\limits_{0}^{\infty}  |F(x)|^{2}\,\frac{\alpha}{x}\, {\mathop{}\!\mathrm{d}} x=\\= \frac{4\pi(9\pi)^{1/3}}{3A^{2/3}} \int\limits_{0}^{\infty} \rho(r)\, r\, {\mathop{}\!\mathrm{d}} r \label{eq:6}
\end{multline}
when the condition
\begin{equation}
\lim_{x\rightarrow \infty} \left[ \rho(r)\,r^2 \right]=0 \label{eq:7}
\end{equation}
is fulfilled. Using Eq.~(\ref{eq:1}) and~(\ref{eq:2}) we obtained~\cite{PhysRevC.109.064621} the explicit relationship of the scaling function with the density distribution:
\begin{multline}
f^\text{QE}(\psi^*)= \frac{4\pi}{A}\int\limits_{0}^{\alpha/(k_{F}|\psi^*|)} \rho(x) \Bigg[ x^2 f_\text{RFG}^\text{QE}[\psi^*(x)]+\\+ \frac{x^3}{3} \frac{\partial f_\text{RFG}^\text{QE}[\psi^*(x)]}{\partial x} \Bigg]\, {\mathop{}\!\mathrm{d}} x. \label{eq:8}
\end{multline}
In our calculations we use the \emph{3pF} parametrization of the $^{12}$C density distribution given in Ref.~\cite{Patt2003}.

\subsection{MEC}

The MEC are characterized as two-body currents that can excite both one-particle one-hole (1p-1h) and two-particle two-hole (2p--2h) states. Numerous investigations into electromagnetic $(e,e')$ processes conducted for low-to-intermediate momentum transfers with MEC in the 1p-1h sector have indicated a slight decrease in the total response at the QE peak, primarily attributed to diagrams that involve the electro excitation of the $\Delta$ resonance; this reduction is approximately balanced by the positive contributions from correlation diagrams, where the virtual photon interacts with a correlated pair of nucleons. It has been found, see Refs.~\cite{VanOrden:1980tg, ALBERICO1984356, PhysRevC.49.2650, DePace:2003spn, Martini:2009uj, PhysRevC.84.055502, Nieves:2011pp, PhysRevD.94.093004, PhysRevC.95.054611}, that the MEC give a significant positive contribution to the cross section, which helps to account for the discrepancy observed in $(e,e')$ processes between theory and experiment in the ``dip'' region between the QE peak and $\Delta$ resonance as well as for the discrepancies between some recent neutrino charge-changing quasielastic (CCQE) measurements (e.g., MiniBooNE, MINERvA, T2K). The presence of \emph{NN} correlation interactions involving the one-nucleon current may result in the excitation of 2p--2h final states, and the interference between these processes and those involving MEC must also be considered. These effects are not explicitly included in our CDFM$_{M^*}$+MEC model. The latter is based on a hybrid description in which the one-particle emission already includes contributions from nuclear ejections due to nuclear correlations, as indicated by the CDFM scaling function.

The process of computing MEC responses requires sophisticated calculations, which involve seven-dimensional integrals of the 2p--2h responses, integration over neutrino energy for the flux-averaged cross-section and possibly an extra averaging over bins in $\cos\theta_\mu$. In the present calculations, unlike those used in our work~\cite{PhysRevC.109.064621} (where the evaluation of the 2p--2h pionic MEC contributions is performed within the RFG model), we utilize a parametrization of the MEC responses which was introduced in Ref.~\cite{PhysRevD.104.113006}. This parametrization introduces semi-empirical formulae that factorize coupling coefficients, form-factors,  phase-space, and the averaged $\Delta$ propagator. The semi-empirical MEC formulae include adjustable coefficients that depend on $q$; these coefficients are listed in Ref.~\cite{PhysRevD.104.113006}. Here, we would like to emphasize that it is certainly more appropriate to use the same effective mass in both contributions (the calculations of the one-body /CDFM-based/ and two-body /semi-empirical MEC parametrization~\cite{PhysRevD.104.113006}/ responses). Thus, our overall calculations still represent a combination of two distinct nuclear models. Therefore, a prospective enhancement of the model involves the evaluation of the 2p--2h MEC contributions within the CDFM$_{M^*}$ model.

In the CDFM$_{M^*}$ model we use as a parameter relativistic effective mass $M^*=m^*_N/m_N=0.8$. A test of the CDFM$_{M^*}$ model for inclusive ($e,e'$) scattering is provided in our papers~\cite{PhysRevC.109.064621, Bxx1}. We demonstrated that the CDFM$_{M^*}$ model description of inclusive ($e,e'$) scattering data is quite acceptable using just one parameter, namely the effective mass $M^*$, which is fixed to $0.8$ in all performed calculations. Before presenting our study of neutrino cross sections, it is essential to first examine the treatment of the $\Delta$ current within the MEC calculation. The $\Delta$ current introduces a range of uncertainties due to its reliance on specific models. Besides the treatment of the propagator, there are also concerns related to form factors and the influence of the nuclear medium. A significant point is the value of the axial form factor value for the excitation of the $\Delta$ at zero momentum transfer, commonly referred to as $C_5^A(0)$. A widely accepted value that has been employed in numerous calculations is $C_5^A(0) = 1.2$. However, in Ref.~\cite{Hernandez:2007qq}, this value was reassessed through a more comprehensive analysis of weak pion emission by the nucleon, resulting in a downward revision to $C_5^A(0) = 0.89$. In this study, we examine how $C_5^A(0)$ contributes to the 2p--2h MEC responses, illustrated as a green band across all figures.

In this study the nuclear medium modifications of the $\Delta$ are not taken into account, rather, the $\Delta$ current is treated as in vacuum. The semiempirical formula presented in Ref.~\cite{PhysRevD.104.113006} is extended to include the case of a $\Delta$ interacting with the mean field. In this case, the $\Delta$ acquires effective mass, $M^{*}_{\Delta}$, and vector energy, $E^{\Delta}_v$ (see Appendix A in~\cite{PhysRevD.104.113006}). An estimate of $M^{*}_{\Delta}=1042$~MeV/$c$ was suggested under the assumption of universal coupling, implying the same vector energy as the nucleon. In Fig. 11 of Ref.~\cite{PhysRevD.104.113006} and Fig.~3 of Ref.~\cite{PhysRevD.108.113006} is shown the comparison of the transverse response functions including or not the $\Delta$ self-energy in the MEC. Our preliminary results related to the double differential flux averaged cross sections, employing the extended semiempirical formula~\cite{PhysRevD.104.113006}, demonstrate that the role of nuclear-medium modifications is negligible for the three experiments analyzed. The study is currently in progress.

\section{Results and discussions\label{sec:results}}

In this section, we present results of the CDFM$_{M^*}$+MEC model for neutrino and antineutrino charged-current quasielastic scattering. An innovative element of our calculations is the introduction, for the first time, of MEC two-nucleon emission derived from the RMF theory for nuclear matter. This approach takes into account dynamical effects in the MEC responses through the relativistic effective mass and vector energy of the nucleon. Our investigation features a comparison with experimental results from the MiniBooNE, T2K, and MINERvA collaborations.

\subsection{MiniBooNE}

\begin{figure*}[p]
\centering\includegraphics[width=.9\textwidth]{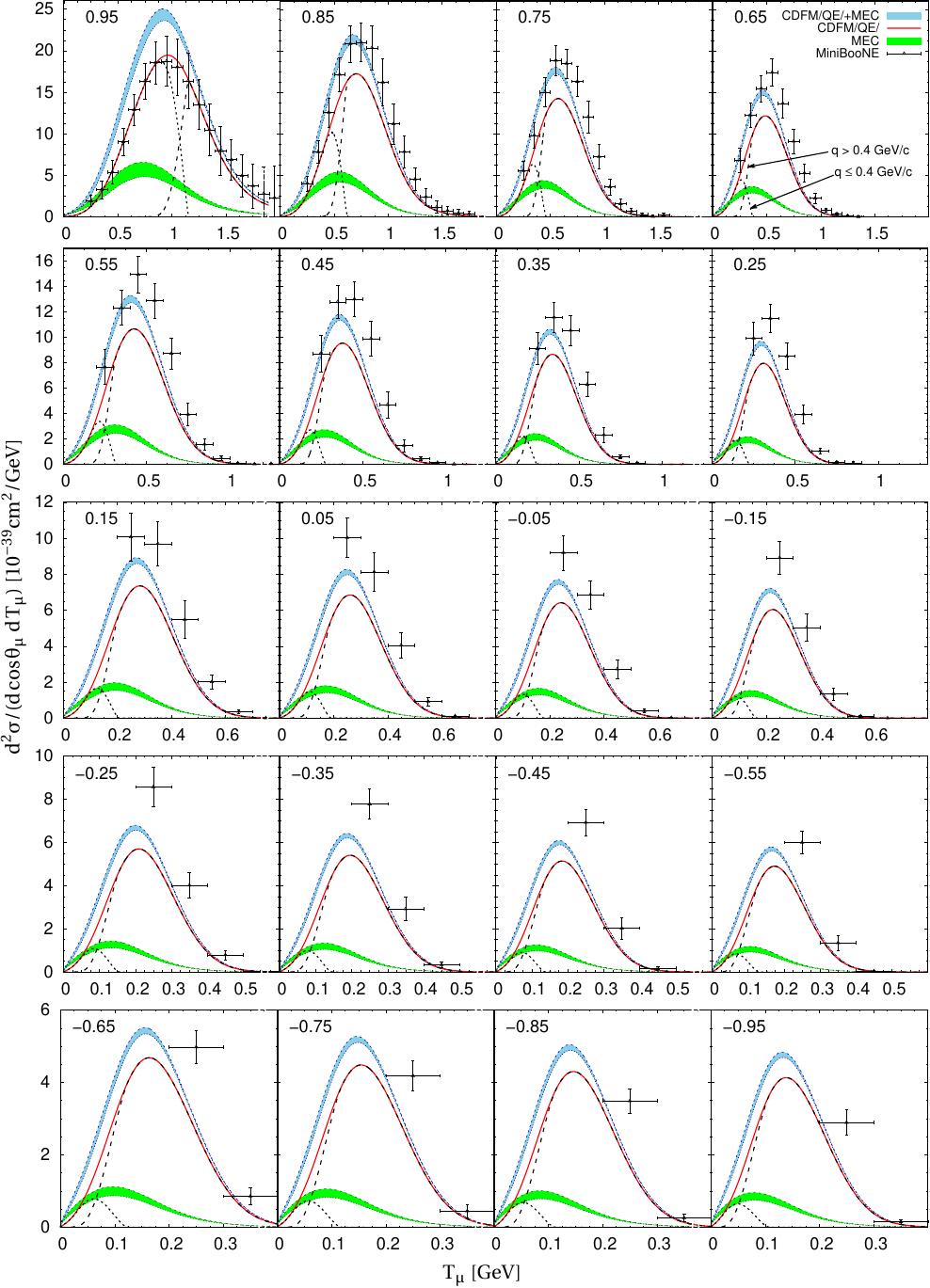}
\caption{(Color online) MiniBooNE flux-folded double differential cross section per target neutron for the $\nu_\mu$ CCQE process on $^{12}$C displayed versus the $\mu^{-}$ kinetic energy $T_\mu$ for various bins of $\cos \theta_\mu$ obtained within the CDFM$_{M^*}$ model including MEC. 2p--2h MEC and QE results are shown separately. The data are from~\cite{PhysRevD.81.092005}.\label{fig:nuMiniBooNE}}
\end{figure*}
\begin{figure*}[t!]
\centering\includegraphics[width=.9\textwidth]{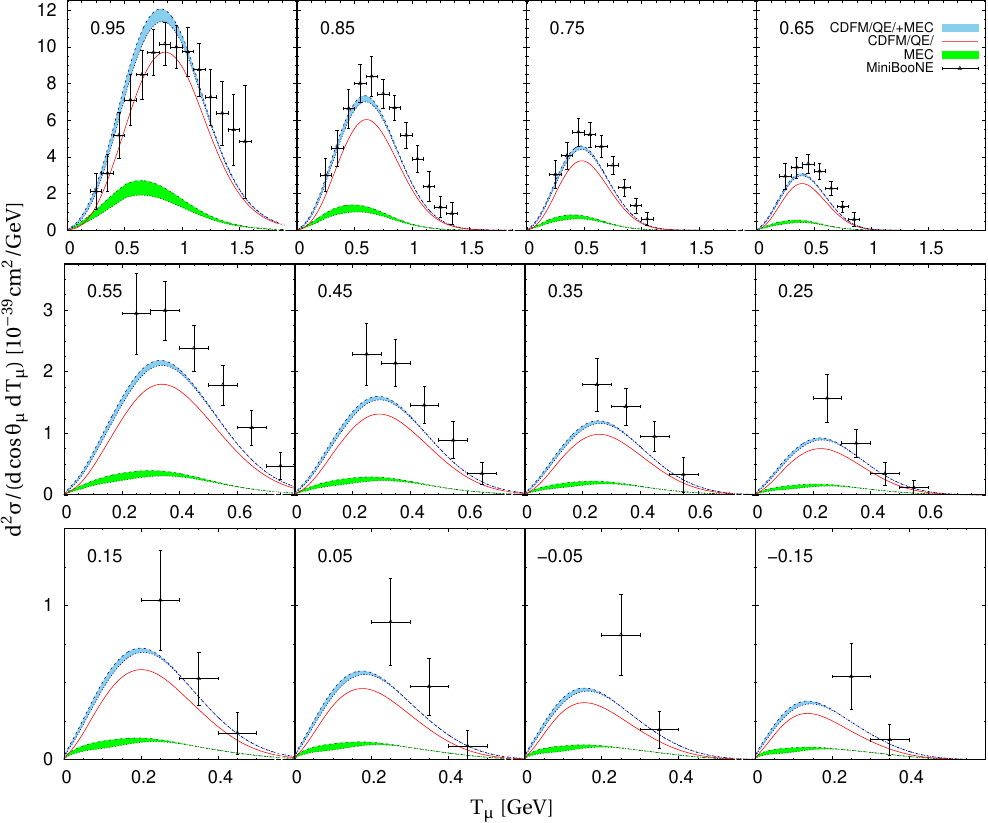}
\caption{(Color online) As for Fig.~\ref{fig:nuMiniBooNE}, but now for the $\overline{\nu}_\mu$ CCQE process on $^{12}$C. The data are from~\cite{miniboone-ant}.\label{fig:anuMiniBooNE}}
\end{figure*}
\begin{figure*}[t!]
\centering\includegraphics[width=.9\textwidth]{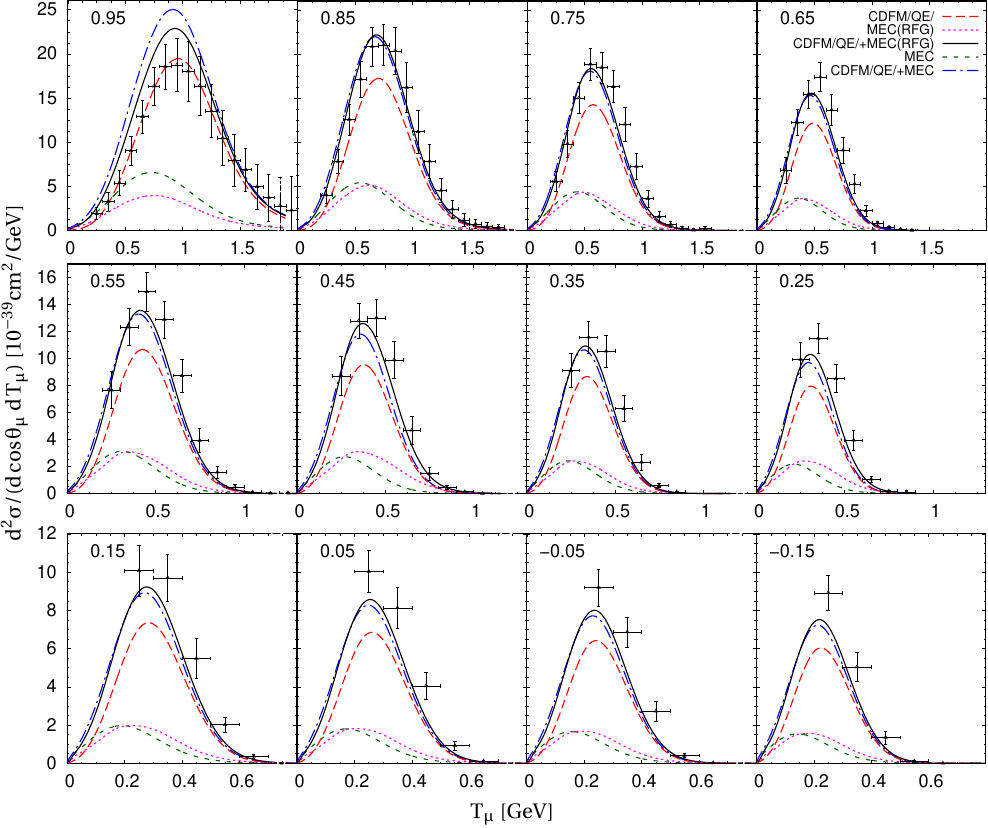}
\caption{(Color online) MiniBooNE flux-folded double differential cross section per target neutron for the $\nu_\mu$ CCQE process on $^{12}$C displayed versus the $\mu^{-}$ kinetic energy $T_\mu$ for various bins of $\cos \theta_\mu$ obtained within the CDFM$_{M^*}$ model including MEC contributions obtained within the RFG model -- MEC(RFG) and using semi-empirical MEC formulae -- MEC. 2p--2h MEC and QE results are shown separately. The data are from~\cite{PhysRevD.81.092005}.\label{fig:nuMiniBooNE1}}
\end{figure*}

Here we present results for the kinematics of the $(\nu_\mu,\mu^-)$ reaction on $^{12}$C as conducted in the MiniBooNE experiment, depicted in  Fig.~\ref{fig:nuMiniBooNE}. In this experiment, the flux-averaged cross-section was measured as a function of the muon energy for fixed values of $\cos\theta_\mu$. The kinematics in terms of $T_\mu$ (muon kinetic energy) and $\cos\theta_\mu$ (angle with respect to the incoming neutrino direction) are averaged over bins. The flux-averaged double differential cross section is computed as
\begin{equation} \label{XSEC}
\frac{d^2\sigma}{dT_\mu d\cos\theta_\mu} = \dfrac{1}{\Phi_{\text{tot}}} \int \Phi(E_\nu) \left.\frac{d^2\sigma}{dT_\mu d\cos\theta_\mu}\right|_{E_\nu}{\mathop{}\!\mathrm{d}}E_\nu,
\end{equation}
where $\Phi(E_\nu)$ is the (anti)neutrino flux, $\left.\frac{d^2\sigma}{dT_\mu d\cos\theta_\mu}\right|_{E_\nu}$ is the cross section for fixed (anti)neutrino energy $E_\nu$ and $\Phi_{\text{tot}}$ is the total integrated $\nu_\mu$($\bar{\nu}_\mu$) flux factor:
\begin{equation}\label{Phitot}
\Phi_{\text{tot}}=\int\Phi(E_\nu){\mathop{}\!\mathrm{d}}E_\nu.
\end{equation}

In Figs.~\ref{fig:nuMiniBooNE} and~\ref{fig:anuMiniBooNE} we show the double differential cross section averaged over the (anti)neutrino energy flux against the kinetic energy of the final muon. The data are taken from the MiniBooNE Collaboration~\cite{PhysRevD.81.092005, miniboone-ant}. The results are displayed in panels representing different bins of $\cos\theta_\mu$ (the width of the experimental bin is $\Delta\cos\theta_\mu=0.1$) and are shown as an average over the bin. In the figures, the QE contribution obtained from the CDFM$_{M^*}$ model is indicated as CDFM/QE/ (solid red line). The 2p--2h MEC contribution is illustrated as a green band, indicating the uncertainties in the evaluation of $C_5^A(0)$ (with the dashed green line representing $C_5^A(0)=0.89$ and the dotted green line indicating $C_5^A(0)=1.2$). The total contribution, labeled CDFM/QE/+MEC (depicted by the blue band), is obtained by summing these two components and is compared with the experimental data from MiniBooNE.

The theoretical results obtained within the framework of the CDFM$_{M^*}$ model, which include both QE and 2p--2h MEC contributions, show a favorable agreement with the data in the majority of the kinematical situations analyzed. Nevertheless, a minor deviation of the data is noted in relation to the theoretical distribution. Also, only at scattering angles close to or above $90^\circ$ can one observe a hint of a difference, however, in these situations, only a small number of data points are available, and they are associated with large uncertainties. It is important to note that the outcomes derived from the CDFM$_{M^*}$ model with $C_5^A(0)=1.2$ align more closely with the data than those obtained with $C_5^A(0)=0.89$.

The CDFM$_{M^*}$ results clearly overestimate the data for extremely forward angles ($\cos\theta_\mu=0.95$), as can be seen in Figs.~\ref{fig:nuMiniBooNE} and~\ref{fig:anuMiniBooNE}. Here, it is important to point out that at small values of the momentum transfer $q$ ($q\leq0.4~\text{GeV}/c$, see Ref.~\cite{PhysRevC.90.035501}) the scaling model is inappropriate due to the failure of the conditions needed for the factorization approximation. At such small momenta, finite-size effects are important and the description of nuclear final states as plane waves is not valid. A possible strategy to improve the scaling approach in this region is to complement it with microscopic calculations (\emph{e.g.}, RPA -- Random Phase Approximation) in the low-$q$ region, which can accurately capture nuclear collective excitations and long-range correlations that are neglected in a purely scaling description. For completeness, the quasielastic results derived from the CDFM$_{M^*}$ model are illustrated by the black dashed line for $q>0.4~\text{GeV}/c$, while those for $q\leq0.4~\text{GeV}/c$ are depicted by the black dotted line in Fig.~\ref{fig:nuMiniBooNE}. The sum of these two contributions gives the quasielastic result obtained within CDFM$_{M*}$ model, represented by a solid red line. One can observe that at very forward angles, the contribution to the QE peak, as well as the left portion of the peak contribution, originates from kinematics where $q\leq0.4~\text{GeV}$. This QE contribution ($q\leq0.4~\text{GeV}$) diminishes as the angle $\theta$ increases.

The contributions from QE and 2p--2h MEC are illustrated separately in Figs.~\ref{fig:nuMiniBooNE} and~\ref{fig:anuMiniBooNE}. It is essential to highlight the significant role that 2p--2h MEC plays to describe accurately the experimental data. The latter accounts for approximately 20--25\% of the total response at its peak. In the case of neutrinos (Figs.~\ref{fig:nuMiniBooNE}), this relative strength remains nearly constant regardless of the scattering angle. Conversely, in the antineutrino case (Fig.~\ref{fig:anuMiniBooNE}), the relative strength of 2p--2h increases for backward scattering angles. This phenomenon arises because the antineutrino cross section exhibits a destructive interference between the $T$ and $T^\prime$ channels, which makes it more responsive to nuclear effects.

For completeness, Fig.~\ref{fig:nuMiniBooNE1} presents a comparison of the theoretical results for the flux-averaged double differential cross section, including 2p--2h MEC contributions obtained using the RFG model -- denoted as MEC(RFG) and those derived from the semi-empirical MEC formulae with $C^A_5 (0) = 1.2$, hereafter referred to as MEC. In our previous studies~\cite{PhysRevC.109.064621, universe11040119}, the 2p--2h MEC contribution was evaluated within the RFG model and is labeled as MEC(RFG) in Fig.~\ref{fig:nuMiniBooNE1}. As shown in the same figure, the 2p--2h MEC results obtained from the two models are in very good agreement in terms of their magnitude, exhibiting only slight relative shift between them. The same behavior is observed in the comparison of results from CDFM/QE/+MEC(RFG) and CDFM/QE/+MEC.

It is worth emphasizing the improvements introduced in the calculations of the MEC 2p--2h responses using the semi-empirical formula adopted in this work, when compared to those based on the RFG model:
\begin{itemize}
\item{} The RMF framework accounts for the effect of the relativistic interactions among nucleons through scalar and vector potentials, resulting in a relativistic effective mass and vector energy.
\item{} The full $\Delta$ propagator is incorporated, including both its real and imaginary parts. The $\Delta$ propagator plays a central role in the MEC framework, as it describes processes in which a nucleon is excited to a $\Delta$-isobar and subsequently decays while interacting with another nucleon. Such processes constitute a key mechanism responsible for two-nucleon emission at high energy transfer.
\end{itemize}

\subsection{T2K}

\begin{figure*}[t!]
\centering\includegraphics[width=1.\textwidth]{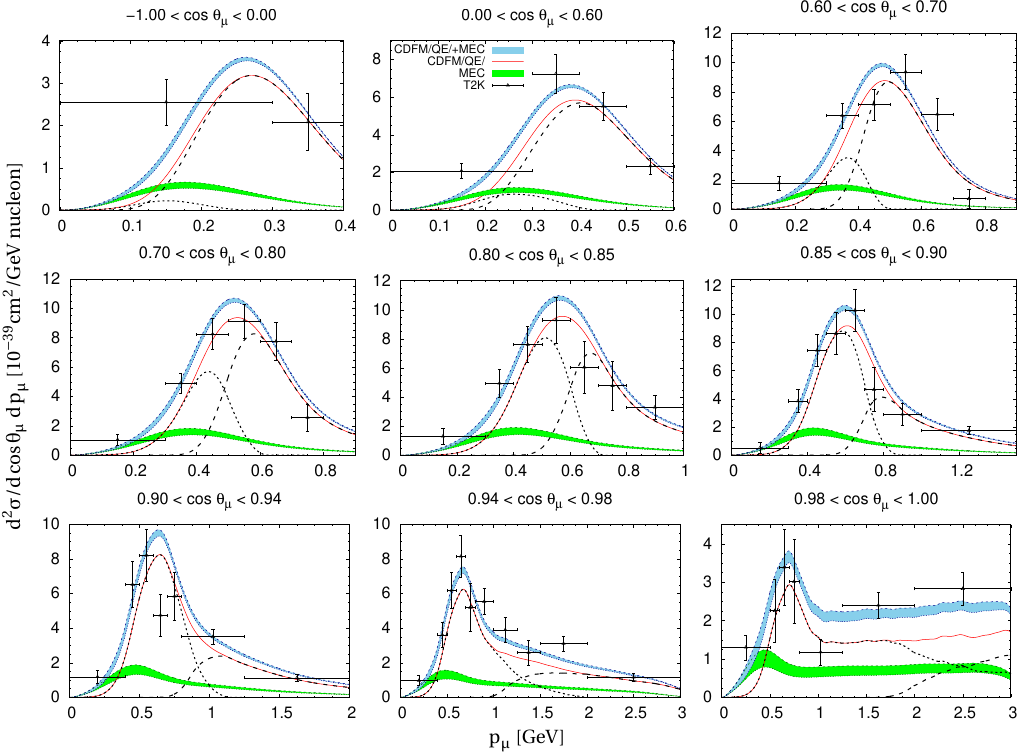}
\caption{(Color online) T2K flux-folded double differential cross section per target nucleon for the $\nu_\mu$ CCQE process on $^{12}$C displayed versus the $\mu^{-}$ momentum $p_\mu$ for various bins of $\cos \theta_\mu$ obtained within the CDFM$_{M^*}$ model including MEC. 2p--2h MEC and QE results are shown separately. The data are from Ref.~\cite{PhysRevD.93.112012}.\label{fig:nuT2K}}
\end{figure*}

\begin{figure*}[t!]
\centering\includegraphics[width=1.\linewidth]{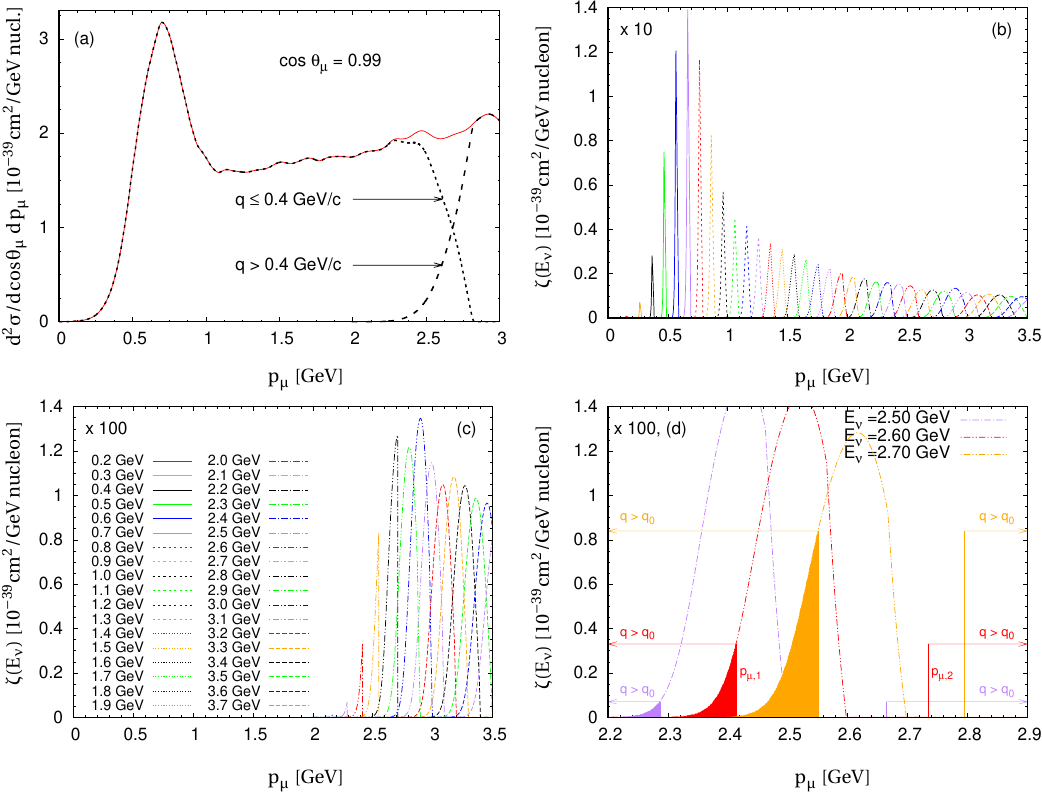}
\caption{(Color online) The quasielastic T2K flux-folded double differential cross section per target nucleon for the $\nu_\mu$--$^{12}$C versus the $\mu^{-}$ momentum $p_\mu$ for $\cos \theta_\mu=0.99$ obtained within the CDFM$_{M^*}$ model [panel (a)]. The panels (b–d) represent the function $\zeta(E_\nu)$ as a function of $p_\mu$ at a scattering angle $\cos \theta = 0.99$, within the framework of the T2K experimental setup (see the text).\label{fig:nuT2K2}}
\end{figure*}

In Fig.~\ref{fig:nuT2K}, we show the flux-folded CC double-differential cross-section for $\nu_\mu$-$^{12}$C scattering derived from the T2K experiment, in comparison with the results obtained within the CDFM$_{M^*}$+MEC model. The graphs are plotted against the muon momentum, and each panel corresponds to a bin in the scattering angle. As in the previous case, we show the results obtained within the CDFM$_{M^*}$ model including MEC and also the separate QE and 2p--2h MEC contributions. As highlighted in~\cite{PhysRevD.94.093004}, the narrower T2K flux, which is sharply peaked at approximately $0.7$~GeV, leads to a reduced contribution from the 2p--2h MEC (around 10\%, with a maximum of 25\% at very forward angles near the QE peak) in contrast to the MiniBooNE results. The dominant contribution to the 2p--2h response is associated with momentum transfers $q \approx 500$~MeV/c, which are less relevant in the context of T2K kinematics. Regarding theoretical predictions, the CDFM$_{M^*}$+MEC model aligns very well with the experimental T2K data for all angles considered.

The results for the quasielstic scattering cross section obtained within the CDFM$_{M^*}$ model, for which $q>0.4~\text{GeV}/c$, are represented by the black dashed line, while those for $q\leq 0.4~\text{GeV}/c $ are plotted by the black dotted line in Fig.~\ref{fig:nuT2K}. As can be seen from Fig.~\ref{fig:nuT2K} in the case of very forward angles the QE contribution comes mainly from $q\leq0.4$~GeV/$c$ in contrast for backward angles where the main contribution comes from $q>0.4$~GeV/$c$.

In Fig.~\ref{fig:nuT2K2} we consider the scattering angle $\cos \theta = 0.99$ and T2K experimental conditions. We investigate the function $\zeta(E_\nu)$ as a function of $p_\mu$, with definition given by
\begin{multline} \label{XSEC_T2K}
\frac{d^2\sigma}{dp_\mu d\cos\theta_\mu} = \dfrac{1}{\Phi_{\text{tot}}} \int \Phi(E_\nu) \left.\frac{d^2\sigma}{dp_\mu d\cos\theta_\mu}\right|_{E_\nu}{\mathop{}\!\mathrm{d}}E_\nu=\\
=\sum\limits_{E_{\nu,\min}}^{E_{\nu,\max}}\dfrac{\Phi(E_\nu)\Delta E_\nu}{\Phi_{\text{tot}}}\left.\frac{d^2\sigma}{dp_\mu d\cos\theta_\mu}\right|_{E_\nu}=\sum\limits_{E_{\nu,\min}}^{E_{\nu,\max}}\zeta(E_\nu),
\end{multline}
where
\begin{equation}\label{eq.zeta}
\zeta(E_\nu) = \dfrac{\Phi(E_\nu)\Delta E_\nu}{\Phi_{\text{tot}}}\left.\frac{d^2\sigma}{dp_\mu d\cos\theta_\mu}\right|_{E_\nu}.
\end{equation}
The results for the quasielstic scattering cross section obtained within the CDFM$_{M^*}$ model using Eq.~(\ref{XSEC_T2K}) are given in Fig.~\ref{fig:nuT2K2}(a) by solid red line. The contributions for which $q>0.4~\text{GeV}/c$ are represented by the black dashed line, while those for $q\leq 0.4~\text{GeV}/c $ are plotted by the black dotted line in Fig.~\ref{fig:nuT2K2}(a). In our calculations, we set $\Delta E_\nu = 1$~MeV. In Fig.~\ref{fig:nuT2K2}(b) are presented the functions $\zeta(E_\nu)$ for several values of neutrino energy, $E_\nu$. As illustrated in Fig.~\ref{fig:nuT2K2}(b), the function $\zeta(E_\nu)$ exhibits a peak behavior. The sum of all peaks yields the flux-averaged cross section at $\cos\theta=0.99$, illustrated by the red solid line in Fig.~\ref{fig:nuT2K2}(a). The functions $\zeta(E_\nu)$ for which $q> 0.4~\text{GeV}/c $ are shown in Fig.~\ref{fig:nuT2K2}(c) and the sum of all peaks gives the black dashed line from Fig.~\ref{fig:nuT2K2}(a). The labels corresponding to the neutrino energy $E_\nu$ are depicted in Fig.~\ref{fig:nuT2K2}(c) and are consistent with those in panels (b) and (d). We would like to emphasize that in panel (b), the results are multiplied by a coefficient of $10$, whereas in panels (c) and (d), the results are multiplied by $100$.

In Fig.~\ref{fig:nuT2K2}(d), we analyze how the peaks $\zeta(E_\nu)$ shown in Fig.~\ref{fig:nuT2K2}(c) are produced based on the inequality $q>q_0$. For our calculations, we fix the value of $q_0=0.4$~GeV/$c$ (see Ref.~\cite{PhysRevC.90.035501}).  The scaling model is not suitable for $q\leq q_0$, due to the inadequacy of the conditions required for the factorization approximation. The momentum transfer $q^2$ can be written as:
\begin{multline}\label{eq.q1}
  q^2=|\mathbf{q}|^2={(\mathbf{p}_\nu-\mathbf{p}_\mu)^2}=
  {E_\nu^2+{p}_\mu^2-2 E_\nu{p}_\mu \cos\theta_\mu},
\end{multline}
where $\mathbf{p}_\nu$ and $\mathbf{p}_\mu$ are the 3-momentum of the incident neutrino and the outgoing muon, respectively. Then the inequality $q^2>q_0^2$ is expressed in the form
\begin{equation}\label{eq.q2}
    {p}_\mu^2-2 E_\nu{p}_\mu \cos\theta_\mu + E_\nu^2- q_0^2>0.
\end{equation}
If the neutrino energy $E_\nu$ is fixed, then Eq.~(\ref{eq.q2}) is a quadratic inequality with respect to the $p_\mu$. Let's consider the discriminant:
\begin{multline}\label{eq.q3}
    {\cal D}= 4E_\nu^2 \cos^2\theta_\mu-4E_\nu^2+4 q_0^2 =4(q_0^2-E_\nu^2 \sin^2\theta_\mu)=\\ 
    =4(q_0-E_\nu \sin\theta_\mu)(q_0+E_\nu \sin\theta_\mu).
\end{multline}
It is clear that if $E_\nu > q_0/\sin\theta_\mu$ (${\cal D}<0$) then the inequality~(\ref{eq.q2}) is fulfilled for any values of $p_\mu$. If $q_0=0.4$~GeV/$c$ the inequality~(\ref{eq.q2}) is satisfied for any values of $p_\mu$ for $E_\nu > q_0/\sin\theta_\mu \approx 2.84$~GeV. This is clearly represented in Fig.~\ref{fig:nuT2K2}(c) for neutrino energies $E_\nu=2.9, 3.0, ...$~GeV. If $E_\nu < q_0/\sin\theta_\mu \approx 2.84$~GeV (${\cal D}>0$) the inequality~(\ref{eq.q2}) is fulfilled outside the interval defined by the roots. In Fig.~\ref{fig:nuT2K2}(d), three cases are examined: $E_\nu = 2.5, ~2.6$, and $2.7$~GeV, detailing how the corresponding peaks for these $E_\nu$ values are obtained. For instance, at $E_\nu = 2.6$~GeV, the values of the two roots are $p_{\mu,1}=2.41$~[GeV] and $p_{\mu,2}=2.73$~[GeV]. Then inequality~(\ref{eq.q2}) is fulfilled within the intervals $p_\mu \in (0,2.41)\cup (2.73,\infty)$~GeV/$c$ as illustrated in Fig.~\ref{fig:nuT2K2}(d) in red color. As shown in Fig.~\ref{fig:nuT2K2}(d), the segment of the function $\zeta(E_\nu)$ at $E_\nu = 2.6$~GeV, where $q>q_0$, is indicated by a red shaded area. This area matches exactly with the peak shown in Fig.~\ref{fig:nuT2K2}(c), which is denoted by the red dash-dot-dot line. Additionally, it can be observed that a reduction in the energy $E_\nu$ leads to a decrease of the peak magnitude of the segment of the function $\zeta(E_\nu)$. This behaviour can be observed also in Fig.~\ref{fig:nuT2K2}(c). At a specific energy $E_0$, the entire peak of the function $\zeta(E_\nu)$ lies between the roots of inequality~(\ref{eq.q2}). Moreover, the function values become negligibly small (effectively zero) outside the interval defined by these roots. The same behavior is observed for the segments of $\zeta(E_\nu)$ corresponding to $q > q_0$. This effect persists for all energies below the energy $E_0$.

It is important to highlight that, although the cross-section contribution from $q<q_0$ plays a significant role in the kinematics illustrated in Fig.~\ref{fig:nuT2K} concerning the QE peak, the CDFM$_{M^*}$+MEC results are in very good agreement with the T2K experimental data.

\subsection{MINERvA}

\begin{figure*}[t!]
\centering\includegraphics[width=.95\textwidth]{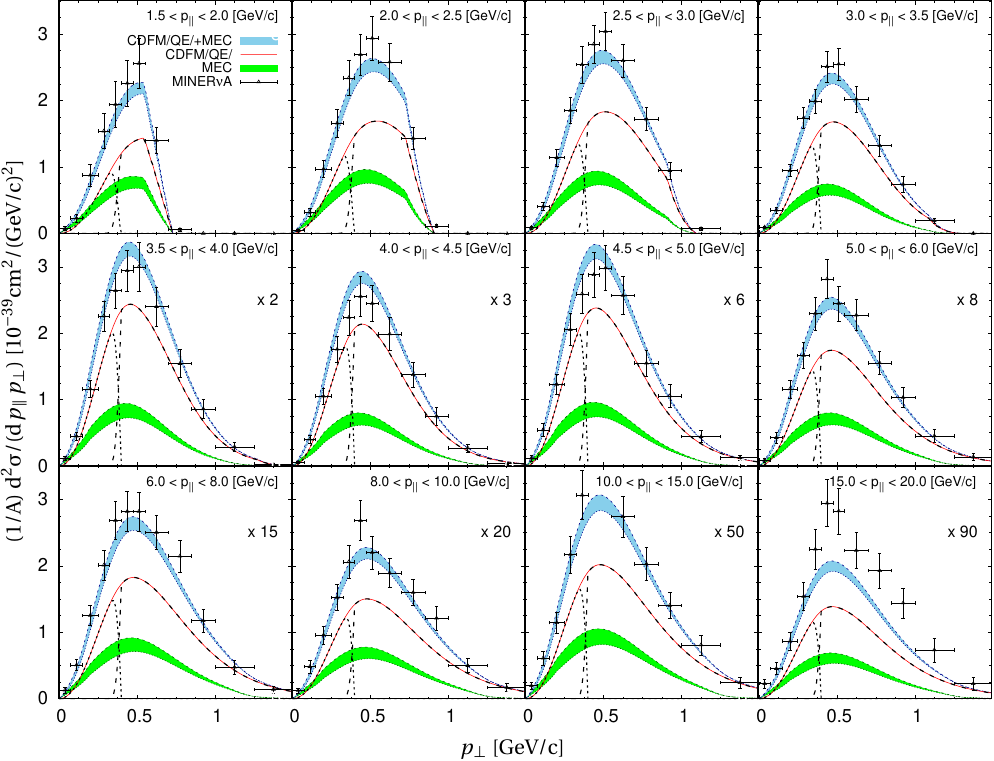}
\caption{(Color online) MINERvA flux-folded double-differential cross section for muon neutrino scattering on hydrocarbon (CH). The QE-like experimental data are from Ref.~\cite{PhysRevD.99.012004}.\label{fig:nuMINERvA}}
\end{figure*}
\begin{figure*}[t!]
\centering\includegraphics[width=.75\textwidth]{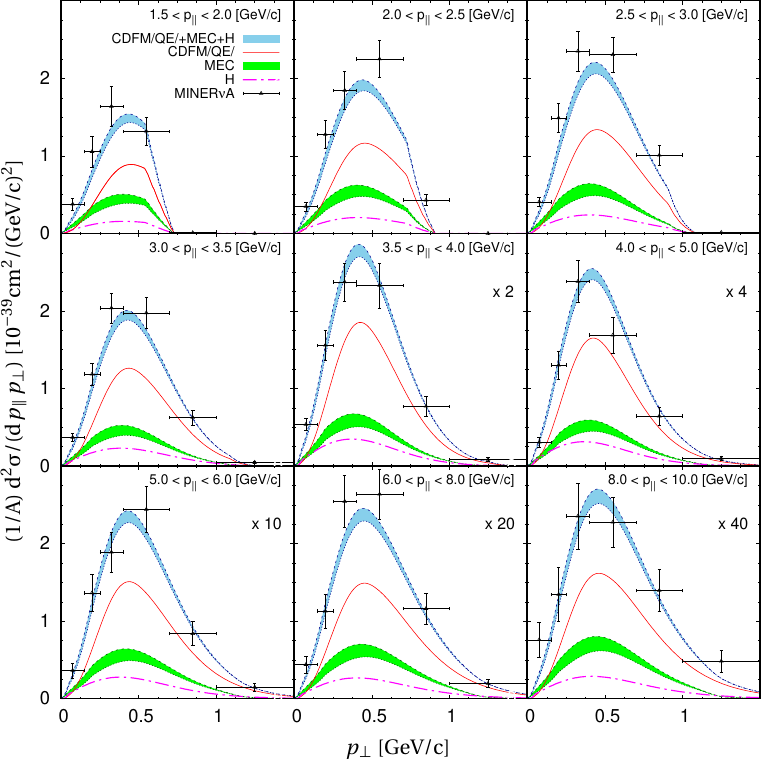}
\caption{(Color online) MINERvA flux-folded double-differential cross section for antineutrino scattering on hydrocarbon (CH). In antineutrino scattering, The single-nucleon contribution for the proton in H is also shown. The QE-like experimental data are from Ref.~\cite{PhysRevD.97.052002}.\label{fig:anuMINERvA}}
\end{figure*}

The data from the MINERvA experiment are presented in Figs.~\ref{fig:nuMINERvA} and~\ref{fig:anuMINERvA} as functions of the measured muon longitudinal and transverse momenta~\cite{PhysRevD.97.052002, PhysRevD.99.012004}. The double-differential cross section is expressed in terms of the longitudinal ($p_\parallel$) and transverse ($p_\perp$) momenta of the scattered muon:
\[
p_\parallel = p_\mu \cos \theta_\mu, \quad p_\perp = p_\mu \sin \theta_\mu.
\]
As a result, the cross section could be represented as:
\begin{equation}
\frac{d^2\sigma}{d p_{\parallel} dp_{\perp}} = \frac{\sin\theta_\mu}{E_\mu} \frac{d^2\sigma}{d E_\mu d\cos\theta_\mu}.
\end{equation}
In Figs.~\ref{fig:nuMINERvA} and~\ref{fig:anuMINERvA} our results, which include both QE derived from the CDFM$_{M^*}$ model and the 2p--2h MEC contributions, are compared with the QE-like data collected using a hydrocarbon target (CH)~\cite{PhysRevD.97.052002, PhysRevD.99.012004} for neutrino and antineutrino scattering, respectively. The figures also illustrate separately the contributions from QE and 2p--2h MEC. It is noteworthy that the 2p--2h MEC plays a crucial role to describe accurately the experimental data, contributing roughly 20–30\% to the total response at its maximum. The theoretical predictions from the CDFM$_{M^*}$ model, which encompass both QE and 2p--2h MEC contributions, align very well with the observed data. Also, as in the previous presented experiments, the quasielastic results obtained from the CDFM$_{M^*}$ model are plotted by the black dashed line for $q>0.4~\text{GeV}/c$, while those for $q\leq0.4~\text{GeV}/c$ are given by the black dotted line in Fig.~\ref{fig:nuMINERvA}.

It is known that the use of an effective mass $M^*=0.8$ to incorporate mean-field effects leads to an excessively large energy shift in the electron-scattering cross section at high momentum transfer, a regime in which the nucleon wave function should approach plane-wave behavior. We emphasize that, motivated by the success of the SuSAv2 model (builds a tradeoff between RMF and Relativistic Plane Wave Impulse Approximation results), an energy-dependent RMF potential (EDRMF) was developed in Ref.~\cite{PhysRevC.100.045501} by modifying the original RMF potential through a blending function that depends on the kinetic energy of the emitted nucleon. The energy-dependent potential preserves the RMF strength and ensures proper orthogonalization for low-momentum nucleons, while becoming softer as the nucleon momentum increases. Ref.~\cite{f7x5-snmz} clearly demonstrates the successful application of the EDRMF model in the MINERvA experiment, which involves a neutrino flux with a higher average energy.

\section{Conclusions\label{sec:conclusions}}

The scaling function $f^\text{QE}(\psi^*)$, constructed within the Coherent Density Fluctuation Model (CDFM$_{M^*}$) using the new scaling variable $\psi^*$ from the interacting relativistic Fermi gas model with scalar and vector interactions, is utilized for the analysis of (anti)neutrino scattering processes. The interacting relativistic Fermi gas model is known to generate a relativistic effective mass for the interacting nucleons. The CDFM$_{M^*}$ model used in our work is based on the $\delta$-function limit of the generator coordinate method and represents a natural extension of the relativistic Fermi gas model to finite nuclei. It is crucial to highlight that CDFM model allows for a quantitative explanation of superscaling in realistic finite systems as it is based on the similar behavior of the high-momentum components of the nucleon momentum distribution from light to heavy nuclei. This is known to be a consequence of the effects of the {\it NN} correlations in nuclei.

The scaling approach of CDFM$_{M^*}$ is used to calculate inclusive (anti)neutrino charge-changing quasielastic differential cross sections and to compare on a consistent basis, the theoretical results with the experimental data from MiniBooNE~\cite{PhysRevD.81.092005, miniboone-ant}, MINERvA~\cite{PhysRevD.99.012004, PhysRevD.97.052002} and T2K~\cite{PhysRevD.93.112012} experiments. The only free parameter in our model is a value of a relativistic effective mass, that is fixed to $m^*_N= 0.8 m_N$ in the calculations. The Fermi momentum $k_F$ is not a free parameter and is calculated within the CDFM$_{M^*}$ model for a given nucleus. In our previous work~\cite{PhysRevC.109.064621}, we used an effective mass of $M^*=1$ for the calculation of 2p--2h meson exchange currents (MEC), which may not be fully consistent with the CDFM$_{M^*}$ calculation, where the dynamical effects of the RMF model are included using $M^*=0.8$. In the present work an alternative parametrization of electroweak 2p--2h MEC responses, computed in the RMF with effective mass $M^*=0.8$, is applied. This parametrization utilizes a semiempirical formula developed in Ref.~\cite{PhysRevD.104.113006}. A test of the CDFM$_{M^*}$ model for inclusive ($e,e'$) scattering is provided in our papers~\cite{PhysRevC.109.064621, Bxx1}.

The CDFM$_{M^*}$ model including MEC is applied to the analyzes of the MiniBooNE (see Figs.~\ref{fig:nuMiniBooNE}, \ref{fig:anuMiniBooNE}, and~\ref{fig:nuMiniBooNE1}), T2K (see Fig.~\ref{fig:nuT2K}), and MINERvA (see Figs.~\ref{fig:nuMINERvA} and~\ref{fig:anuMINERvA}) experiments. The theoretical results obtained within the CDFM$_{M^*}$ model including both QE and 2p--2h MEC are in very good agreement with the data from the analyzed experiments in most of the kinematical situations considered in the present work. The analysis indicates that the impact of the 2p--2h MEC effects can reach approximately 20--30\% when compared to the pure QE responses in the MiniBooNE, around 10\% in the T2K (only for very forward angles up to 25\%), and between 30--50\% in the MINERvA experiments, depending on the kinematics considered and the value of $C_5^A(0)$ used in the calculations. In this study, we also analyze the influence of the uncertainty in the axial form factor value concerning the excitation of the $\Delta$ at zero momentum transfer, $C^A_5 (0)$. The results for neutrino and antineutrino CCQE differential cross sections obtained through the CDFM$_{M^*}$+MEC model demonstrate that results with $C^A_5 (0)=1.2$ are more consistent with the experimental data.

In this work we present a comprehensive analysis of the quasielastic results within the CDFM$_{M^*}$ model for both $q > 0.4$~GeV/$c$ and $q \leq 0.4$~GeV/$c$ across all neutrino kinematics in the three experiments considered. As discussed in Ref.~\cite{PhysRevC.90.035501}, the scaling model is less reliable at low $q$ ($q<0.4$~GeV/$c$) due to the breakdown of the factorization conditions. Nevertheless, although the contribution from $q<0.4$~GeV is significant in certain  kinematics (Figs.~\ref{fig:nuMiniBooNE}, \ref{fig:nuT2K}, and~\ref{fig:nuMINERvA}) near the quasielastic peak, the CDFM$_{M^*}$+MEC provides predictions in good agreement with the experimental data. The results of the present work allow us to conclude that our approach is capable to be applied to analyses of (anti)neutrino scattering on the $^{12}$C nucleus. Moreover, our approach is easily extendable to heavier nuclei, particularly in view of the recent MicroBooNE experimental results regarding (anti)neutrino interactions with the $^{40}$Ar nucleus.

As a general conclusion we note that in the present work we consider the important problem of the neutrino-nuclei interaction that is related to the fundamental question of the neutrino oscillations. Our approach uses the CDFM$_{M^*}$ model that allows one to transform the nuclear matter quantities to the corresponding ones in finite nuclei. The model includes the possibility to account for the nucleon-nucleon correlations in nuclei whose effects are responsible for the superscaling that is one of the most important phenomena of the lepton interactions with nuclei.

\begin{acknowledgements}
This work was partially supported by the Bulgarian National Science Fund under contract No.~KP-06-N98/3.
\end{acknowledgements}

\bibliography{biblio}

\end{document}